\documentclass
[prl,reprint,showpacs,showkeys,final,nobibnotes,nofootinbib,onecolumn,11pt]{revtex4}%
\usepackage{amsmath}
\usepackage{amsfonts}
\usepackage{amssymb}
\usepackage{graphicx}
\usepackage{rotating}%
\providecommand{\U}[1]{\protect\rule{.1in}{.1in}}

\begin{document}
\title{Least square based method for obtaining one-particle spectral functions from
temperature Green functions}
\author{Jun Liu}
\email{junliu@ameslab.gov}
\affiliation{Ames Laboratory-US DOE and Department of Physics and Astronomy, Iowa State
University, Ames, IA, 50010}
\date{September, 24th, 2011}

\begin{abstract}
A least square based fitting scheme is proposed to extract an optimal one
particle spectral function out of any one-particle temperature Green function.
It uses the existing non-negative least square(NNLS) fit algorithm to do the
fit, and the Tikhonov regularization to help with possible numerical singular
behaviors. By flexibly adding delta peaks to represent very sharp features of
the target spectrum, this scheme guarantees a global minimization of the
fitted residue. The performance of this scheme is manifested with diverse
physical examples. The proposed scheme is shown to be comparable in
performance to the standard Pade analytic continuation scheme.

\end{abstract}
\keywords{one-particle spectral function, temperature Green function, non-negative least
square fit (NNLS), Tikhonov regularization, Pade analytic continuation, global minimization}
\pacs{02.60.Gf; 02.60.Nm; 64.60.De}
\maketitle

An inverse problem tries to make an inference on a cause which is related to a
response experimentally accessible by some well-established equation. Usually
the relation can be either an integral or a differential equation, which
cannot be solved analytically but be treated with utmost numerical
care\cite{IntroductionToTheTheoryOfInverseProblems}. Although being widely
involved in different physics
fields\cite{Inverse_optical_PhysRevB.73.184507,SVD_phonon_fit_PhysRevB.71.104529,Baym_JMathPhys.2.232}
as well as other areas\cite{AnIntroductionToElectromagneticInverseScattering},
the inverse problem has largely been an open issue to be fully addressed till
now. In this paper, we will focus on a typical inverse problem, the notorious
analytic continuation problem, and present a scheme which is comparable to the
Pade scheme in their performances.

The analytic continuation problem concerns inferring a spectral function from
some input temperature Green function. A temperature Green function,
$\mathcal{G}\left(  i\omega_{n}\right)  ,$ is related to a one-particle
spectral function, $A\left(  \Omega\right)  ,$ through an integral equation
expressed in Eq. \ref{0-1} as
\begin{equation}
\mathcal{G}\left(  i\omega_{n}\right)  =\int_{-\infty}^{\infty}\frac{A\left(
\Omega\right)  }{i\omega_{n}-\Omega}\frac{d\Omega}{2\pi} \label{0-1}%
\end{equation}
where $\omega_{n}$ can be either fermionic or bosonic Matsubara
frequencies\cite{QuanTheOfMBS}. Baym and Mermin\cite{Baym_JMathPhys.2.232}
showed unique existence of $A\left(  \Omega\right)  $ given $\mathcal{G}%
\left(  i\omega_{n}\right)  $ at all Matsubara frequency points. Indeed, if
the analytic expression is known for $\mathcal{G}\left(  i\omega_{n}\right)
,$ the spectral function can be directly determined via Eq. \ref{0-2}%
\begin{equation}
A\left(  \omega\right)  =-\frac{1}{\pi}\lim_{\eta\rightarrow0^{+}%
}\operatorname{Im}\mathcal{G}\left(  \omega+i\eta\right)  \label{0-2}%
\end{equation}
with the proper analytic continuation to the real axis by replacing
$i\omega_{n}$ with $\omega+i\eta$. However, an analytic expression of
$\mathcal{G}\left(  i\omega_{n}\right)  $ is usually not available as it is
most often evaluated numerically through Feynmann diagrammatic techniques or
other numerical
schemes\cite{flex_AnnPhys.193.206,1st_QMC_PhysRevB.26.5033,QuanTheOfMBS}. In
these cases, one might have to invert Eq. \ref{0-1} directly to get an
estimation for $A\left(  \Omega\right)  ,$ where the unique existence property
might be blurred due to numerical errors.

Two types of algorithms are widely used to numerically carry out the analytic
continuation depending on different natures of $\mathcal{G}\left(  i\omega
_{n}\right)  $. In case the input data has very high numerical accuracy, the
widely used method is the Pade scheme\cite{Vidberg_JLowTempPhys.29.179}, where
the $i\omega_{n}$ dependence of the temperature Green function is approximated
with a continued fraction polynomial with higher order expansion terms
truncated\cite{EssPadeAppr}. By construction, the Pade scheme rigorously
reproduces select $\mathcal{G}\left(  i\omega_{n}\right)  $ points to fulfill
the spirit of an analytic continuation. It can give very good results, but
sporadically it could also fail by returning negative spectrum or fake sharp
peaks in the spectrum. It would thus be ideal if the Pade scheme could be
improved free of the above mentioned issues. However, we may argue here that
this is not possible. Actually, it is not guaranteed that a valid spectral
function is always attainable with an analytic continuation scheme designed to
completely reproduce a temperature Green function of finite precision. The
logic is the following. Given any temperature Green function, $\mathcal{G},$
we can construct a new temperature Green function, $\mathcal{\tilde{G}},$
defined on a subset of $\mathcal{G}$ as, say,
\begin{equation}
\mathcal{\tilde{G}}\left(  i\tilde{\omega}_{n}\right)  =\mathcal{G}\left(
i\omega_{3n+1}\right)
\end{equation}
for $n\in\mathcal{N}.$ $\mathcal{\tilde{G}}$ corresponds to a virtual system
whose temperature is increased three times that of the original system in
which $\mathcal{G}$ is defined. By construction, $\mathcal{\tilde{G}}$ shares
the same spectral function with $\mathcal{G}$, a direct outcome of the theorem
by Baym and Mermin\cite{comment3_SameSpectr4Gt&G}. Now, let's add an arbitrary
perturbation to the $i\omega_{0}$ component of $\mathcal{G}$ and thereby
define a second temperature Green function, $\mathcal{G}^{\prime}.$ We
immediately see that $\mathcal{G}$ and $\mathcal{G}^{\prime}$ cannot share the
same spectral function. Actually there exists no valid spectral function which
gives out $\mathcal{G}^{\prime},$ as this spectral function should also give
out $\mathcal{\tilde{G}}$ by noting that $\mathcal{\tilde{G}}$ is an infinite
size subset of $\mathcal{G}^{\prime}$ as well. On the other hand, however, the
Pade type schemes can always be applied to $\mathcal{G}^{\prime}$ and give
rise to a "spectral function", which cannot be rigorously positive definite as
is argued above. We believe that the fact that the Pade scheme is excessively
sensitive to numerical round-off errors supports the claim made here.

When the input Green function has a known covariance structure, e.g. data
coming from Quantum Monte Carlo simulations\cite{1st_QMC_PhysRevB.26.5033},
the analytic continuation has been carried out in a totally different way from
the Pade scheme and its alike. Data matching is now not the best thing to do
for an analytic continuation due to data error, but instead, models balancing
between the input data and the known covariance structure are what people have
been looking for. The widely used methods include the Maximum Entropy
Method(MEM), the stochastic method and their
extensions\cite{Jarrell_MEM_PhysRevB.44.5347, Sebastian_SA_PhysRevE.81.056701,
Sandvik_SA_PhysRevB.57.10287, Syljuasen_ASM_PhysRevB.78.174429}, where
Baysian-type arguments are heavily used. Other methods were tried out as
well\cite{early_QMC_chi2_fit_PhysRevB.34.4744,Vitali_GA_PhysRevB.82.174510,
OptStochReg_Arxiv.cond-mat/0612233v3}. All these research efforts seem to
leave one a plausible impression that there is a gap between treating data
with or without a known covariance structure when doing an analytic
continuation. We believe that this is not necessarily the case if we can find
a robust and accurate analytic continuation scheme to treat data with high precision.

Generally speaking, an inverse problem cannot be fully solved numerically
unless the input data and the computer system used for the calculation have
extremely high precision, as is the case for the analytic continuation
problem\cite{Beach_PhysRevB.61.5147}. This, together with several practical
considerations, suggests some desirable features of a new scheme for doing
analytic continuation.

\begin{enumerate}
\item The scheme should be robust against numerical round-off errors in the
input data.

\item The constraints on the solution should be fully observed to exclude
unphysical solutions and to avoid multiple solutions

\item The final solution should be reliable and repeatable, which mainly
implies that the scheme has a unique and easily accessible global minimum.

\item The equation establishing the inverse problem should be best preserved
when it is transformed into a convenient form for subsequent treatment.

\item Every input data point should be fitted to the statistically highest
precision allowed by a computer.
\end{enumerate}

The first point concerns about the scheme as a whole to be practically
convenient. The next two points help to render a solution of less
arbitrariness. The last two points aim at generating a most accurate solution.
The Pade scheme only fulfills conditions (3) and (5). Its sporadic failure
suffered from numerical error prevents it from having even wider applications.
The Bayesian-type methods don't seem to satisfy condition (4) and (5) well,
thus results coming out of these methods need to be justified by being
compared against relevant experiments. We believe that a carefully designed
least square fitting scheme can best compromise among the above conditions.
There had been some trials with least square based
methods\cite{early_QMC_chi2_fit_PhysRevB.34.4744,SVD_spectral_fit_PhysRevLett.75.517,
Beach_PhysRevB.61.5147}. Former attempts along this line of thought were not
carried out carefully enough on approximating the functional
form\cite{early_QMC_chi2_fit_PhysRevB.34.4744,
SVD_spectral_fit_PhysRevLett.75.517} and on preserving
nonnegativeness\cite{SVD_phonon_fit_PhysRevB.71.104529} of the spectrum. Here
we introduce a novel least square based method to fully satisfy all the above
mentioned conditions with a slight exception on condition (4).

Very generally, we can assume $A\left(  \Omega\right)  $ to be parametrized
as
\begin{equation}
A\left(  \Omega\right)  =\tilde{A}\left(  \Omega\right)  +\sum_{l=1}^{L}%
I_{l}\delta\left(  \Omega-\tilde{\Omega}_{l}\right)  \label{1-2}%
\end{equation}
where $\tilde{A}\left(  \Omega\right)  $ is the continuous part of the
spectrum and $\delta\left(  \Omega-\tilde{\Omega}_{l}\right)  $ represents
very sharp quasiparticle peaks which could exist in the system. $I_{l}$ is the
spectrum weight of a quasiparticle peak located at $\tilde{\Omega}_{l}$. We
usually don't know the functional form of $\tilde{A}\left(  \Omega\right)  $.
But as a very crude yet later proven to be very efficient step, we approximate
$\tilde{A}\left(  \Omega\right)  $ with joined linear segments within the
frequency range of interest\cite{comment1_linear_appr}. Denoting $\left\{
\Omega_{m}\right\}  $ with $m=1,2,\ldots,M$ the pre-chosen frequency grid, and
$\left\{  A_{m}\right\}  $ the unknown magnitudes on the corresponding
frequency points, we have%
\begin{equation}
\tilde{A}\left(  \Omega\right)  =\frac{A_{m+1}-A_{m}}{\Omega_{m+1}-\Omega_{m}%
}\left(  \Omega-\Omega_{m}\right)  +A_{m} \label{1-3}%
\end{equation}
for $\forall\Omega\in\left[  \Omega_{m},\Omega_{m+1}\right]  .$ If we feed Eq.
\ref{1-3} and Eq. \ref{1-2} back to Eq. \ref{0-1} and integrate over each
segment, we obtain a linear equation in terms of $A_{i}$ and $I_{i}$,%
\begin{align}
\mathcal{G}\left(  i\omega_{n}\right)   &  =\sum_{m=1}^{M-1}\left[
\frac{i\omega_{n}-\Omega_{m-1}}{\Omega_{m}-\Omega_{m-1}}\ln\left(
\frac{i\omega_{n}-\Omega_{m-1}}{i\omega_{n}-\Omega_{m}}\right)  +\frac
{\Omega_{m+1}-i\omega_{n}}{\Omega_{m+1}-\Omega_{m}}\ln\left(  \frac
{i\omega_{n}-\Omega_{m}}{i\omega_{n}-\Omega_{m+1}}\right)  \right]
A_{m}\nonumber\\
&  +\sum_{l=1}^{L}\frac{I_{l}}{i\omega_{n}-\tilde{\Omega}_{l}} \label{1-4}%
\end{align}
Note any integrations incurred in the calculation are carried out
analytically. This marks the first difference between our scheme and the
previous existing efforts\cite{SVD_phonon_fit_PhysRevB.71.104529,
SVD_spectral_fit_PhysRevLett.75.517}, where discretization error was
introduced as the continuous frequency axis was discretized into a frequency
grid set up on $\left\{  \Omega_{m}\right\}  $\cite{comment5_PrevLSF}. One
more relation can be added into Eq. \ref{1-4} by noticing that $A\left(
\Omega\right)  $ needs to be normalized to reflect particle
conservation\cite{comment2_normal}. Meanwhile, the positive definiteness of a
spectrum for fermions becomes equivalent to the inequality constraints%
\begin{equation}
A_{m},I_{l}\geq0 \label{1-5}%
\end{equation}
under the current construction, which should be rigorously fulfilled. For
bosons, the above condition should be slightly modified. Eq. \ref{1-4},
together with the non-negative constraints in Eq. \ref{1-5}, is readily
treated by the well-studied non-negative least square fitting
algorithm(NNLS)\cite{SolvingLeastSquaresProblems} if the coefficient matrix
formed by the linear coefficients in Eq. \ref{1-4} is well-behaved, which is
however seldomly the case. To resolve this, we resort to the Tikhonov
regularization scheme\cite{NumericalMethodsforLeastSquaresProblems}. The
Tikhonov matrix is chosen to be the identity matrix multiplying some cutoff
parameter, $\lambda,$ which is set to be about two orders of magnitude higher
than the accuracy of the input data, at $10^{-14}$ if not mentioned otherwise
in this paper. By doing so, noises are expected to be safely filtered out and
only a healthy spectrum is left behind. Note that usually the accuracy of
input data is not known \textit{a prior}. How to reasonably estimate this
quantity is suggested in the next paragraph.

In all, given a temperature Green function defined on $N$ Matsbuara frequency
points, the Least Square Fitting (LSF) scheme introduced in this paper solves
the linear system of equations,
\begin{equation}
\left(
\begin{array}
[c]{c}%
\bar{\Phi}\\
\lambda I_{\left(  M+L-1\right)  \times\left(  M+L-1\right)  }%
\end{array}
\right)  \bar{A}=\left(
\begin{array}
[c]{c}%
\bar{\mathcal{G}}\\
\bar{0}_{\left(  M+L-1\right)  \times1}%
\end{array}
\right)  \label{1-6}%
\end{equation}
for the spectral function. Here, $\bar{\Phi}$ is a $\left(  M+L-1\right)
\times N$ matrix with the matrix elements defined through Eq. \ref{1-4} as
\begin{equation}
\bar{\Phi}_{m,n}=\left\{
\begin{array}
[c]{cc}%
\dfrac{i\omega_{n}-\Omega_{m-1}}{\Omega_{m}-\Omega_{m-1}}\ln\left(
\dfrac{i\omega_{n}-\Omega_{m-1}}{i\omega_{n}-\Omega_{m}}\right)
+\dfrac{\Omega_{m+1}-i\omega_{n}}{\Omega_{m+1}-\Omega_{m}}\ln\left(
\dfrac{i\omega_{n}-\Omega_{m}}{i\omega_{n}-\Omega_{m+1}}\right)  & \text{if
}m\leq M-1\\
\dfrac{1}{i\omega_{n}-\tilde{\Omega}_{m-\left(  M-1\right)  }} & \text{if
}m>M-1
\end{array}
\right.
\end{equation}
$\bar{A}$ is the unknown spectral magnitudes defined as
\begin{equation}
\bar{A}_{m}=\left\{
\begin{array}
[c]{cc}%
A_{m} & \text{if }m\leq M-1\\
I_{m-\left(  M-1\right)  } & \text{if }m>M-1
\end{array}
\right.
\end{equation}
and $\bar{\mathcal{G}},$ the input temperature Green function, is simply
defined as
\begin{equation}
\bar{\mathcal{G}}_{n}=\mathcal{G}\left(  i\omega_{n}\right)
\end{equation}
As is usually defined, $I_{\left(  M+L-1\right)  \times\left(  M+L-1\right)
}$ and $\bar{0}_{\left(  M+L-1\right)  \times1}$ denote an identity matrix and
a zero vector respectively.

\begin{figure}[ptb]%
\centering
\includegraphics[
height=3.0303in,
width=4.9753in
]%
{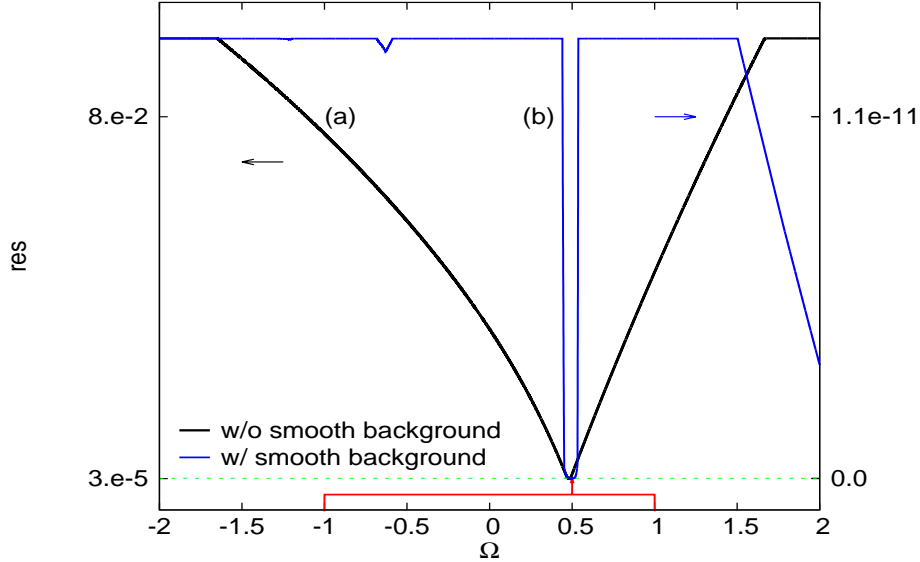}%
\caption{Running residue defined as difference between a fitted and an input
Green function for a model system, which has a flat spectrum with a delta peak
located around $\Omega\approx0.5,$ as illustrated in red at the bottom. Curve
(a) is calculated by scanning the whole frequency range with just one delta
peak, while curve (b) is calculated with a delta peak as well as a continuous
background spectrum $\tilde{A}\left(  \Omega\right)  .$ The x axis denotes the
scanning position of the delta peak. The y axis denotes the residue defined in
Eq. \ref{1-7} when the scanning delta peak reaches a specific frequency
$\Omega$. Note that in the plot the two residues are scaled to be comparable
to each other. }%
\label{fig1}%
\end{figure}

A crucial piece is missing from the above formalism, that is, we still don't
know how to determine the number and location of quasiparticle peaks to be
included in the fit if a continuous spectrum alone cannot give a successful
fit to the input Green function. Note that the residue,%
\begin{equation}
res=\left\Vert \bar{\mathcal{G}}-\bar{\Phi}\bar{A}\right\Vert \label{1-7}%
\end{equation}
returned by Eq. \ref{1-6}, seems to be "screened" by the continuous spectrum,
as is illustrated in Fig \ref{fig1}. This shows that the residue is not
sensitive to the location of a tentative delta peak unless it is very close to
a true one.
To be more specific, we can see that a fit with help of a continuous spectrum
gives a very abrupt change in its residue around the true delta peak location,
and is much more confined in width than a fit without such a continuous
background. The width of the residue drop is approximately $0.09$ when the
input data is fitted by a delta peak with a continuous background. This gives
the "highest precision" in determining the location of a delta peak with the
current fitting scheme, where any delta peaks within this distance cannot be
distinugished. The abrupt drop of residue and its local nature implies two
things. Firstly this makes it necessary to carry out a full scan within the
interested frequency range to find out all possible quasiparticle locations by
checking obvious residue drops, which is not too time-consuming for a one
dimensional scan. Secondly, this actually helps the minimization process to
reach the global minimum as variables are not strongly dependent on each other
any more. If the returned residue is not small enough, one can always repeat
the procedure by adding in more peaks until no peaks should be added.
Dynamically including more delta peaks into the NNLS fit distinguishes this
scheme from all previous trials. As the global minimum can be ensured in this
way, the precision of a set of input data can be readily deduced by the best
possible fit to the input data with this scheme.

When a fit is to be tested against some known results, the delta peaks need to
be transformed such that a direct comparison against the true spectral
function is possible. To do that, these delta peaks need to be expanded with
help of Eq. \ref{0-2} as
\begin{equation}
\tilde{I}_{i}\left(  \Omega\right)  =\frac{1}{\pi}\frac{\eta I_{i}}{\left(
\Omega-\Omega_{i}\right)  ^{2}+\eta^{2}} \label{1-3a}%
\end{equation}
We choose $\eta=0.01$ consistently in this work to generate a finite width of
a peak and an infinitesmally positive shift to rotate the Green function
coming out of the Pade scheme onto the real axis.

\begin{figure}[ptb]%
\centering
\includegraphics[
trim=0.000000in 0.016199in 0.000000in -0.016199in,
height=3.9176in,
width=5.0272in
]%
{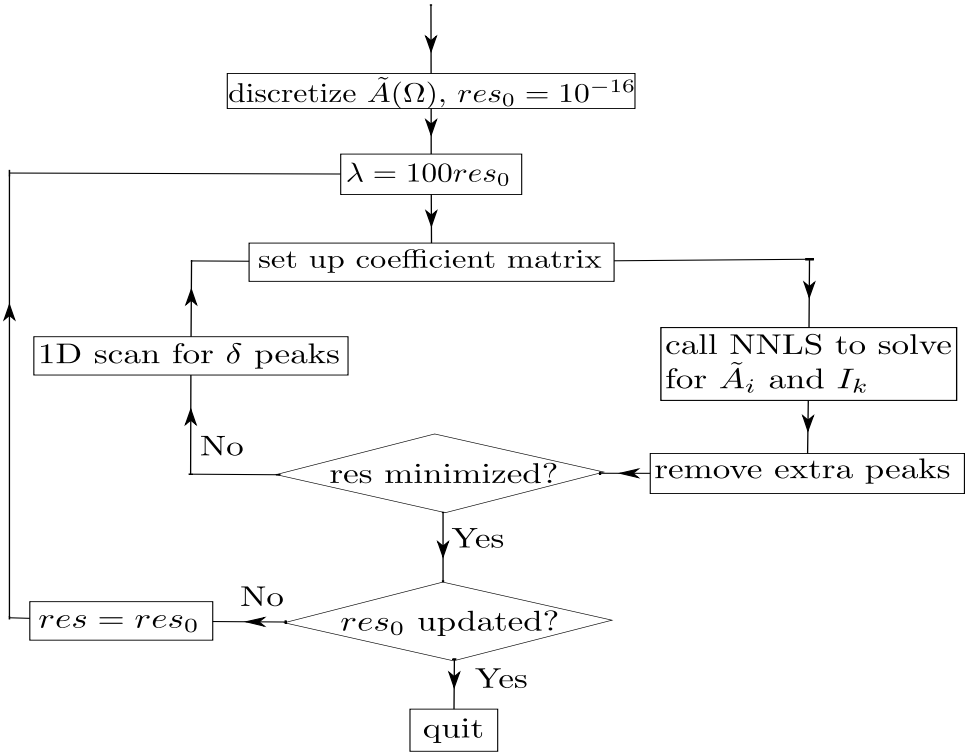}%
\caption{Flow chart of the Least Square Fitting (LSF) scheme. The whole scheme
is made up of two loops. The inner loop obtains a best fit to the input Green
function for a prefixed error of the input Green function, denoted as
res$_{0}.$ The resulting best fitted error, res$,$ is then used to update
res$_{0}$ for the outer loop to correct the Tikhonov regularization
coefficient $\lambda,$ which controls error propagating from the data to the
fitted spectrum in the inner loop. Usually res$_{0}$ just needs to be updated
once before the whole fit is done.}%
\label{fig1.1}%
\end{figure}

The details above outline the general idea on how to carry out the LSF scheme,
as illustrated in Fig. \ref{fig1.1}.%
The portion of the fit with a continuous background is easily implemented, but
the portion related to isolated delta peaks becomes very much involved in
dealing with various practical concerns, say, how to remove extra tentative
peaks added with an intermediate fit, how to push the fit to the extremum of
machine precision, etc. After all these issues are carefully addressed, we
then look at several examples involving fermions to study the performance of
this scheme, using the Pade analytic continuation
scheme\cite{Vidberg_JLowTempPhys.29.179} based on the continued fraction
polynomial as a benchmark comparison. For each example, the analytic
continuation is carried out with a fit on the lowest $128$ positive Matsubara
frequency points for four different temperatures. To make things simpler, we
will assume the interested frequency region is known to us. The region is
evenly divided to set up the linear system as in Eq. \ref{1-6}, with each
interval at least twice the width of a typical residue drop if not mentioned otherwise.

Firstly, two toy model problems are considered whose temperature Green
functions are set up through Eq. \ref{0-1} with artificially constructed
spectral functions. The first example considers a step-wise spectral function.
We use this example to study the performance of the fitting scheme when a
system has a discontinuous spectral function. The second example mimics the
Mott-insulating transition under intermediate Hubbard U, which brings up some
quasiparticle peaks near the Fermi energy and splits a conduction band into an
upper and a lower subband\cite{DMFT_RevModPhys.68.13}. This example mainly
stresses on how well the current scheme smoothes through an extremely volatile
conduction band, and shows how it could accurately reproduce the delta peaks.
With a finite precision for the input data, it is impossible to reproduce
every detail of the spectrum in an inverse problem. The cutoff parameter,
$\lambda,$ used in the Tikhonov scheme, plays the role of controlling how
accurately the true spectrum function can be reproduced.%

\begin{figure}[ptb]%
\centering
\includegraphics[
height=3.9963in,
width=5.706in
]%
{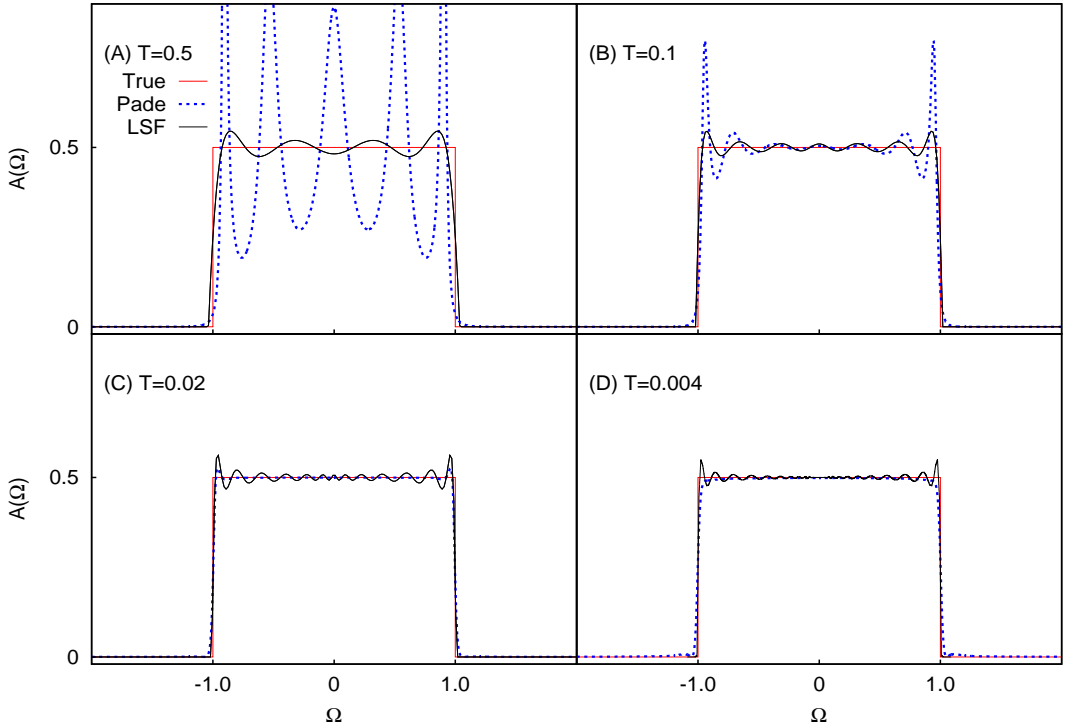}%
\caption{Comparison of the true spectrum function against those returned from
the Pade and the LSF schemes for the same step-wise spectrum under different
temperatures. For the LSF scheme, evenly divided frequency partition on x axis
does not improve fitting quality noticeably beyond a critical number of
intervals. It is not very stable for the Pade scheme to work for the current
example. Specifically, it requires the number of Matsubara frequency points to
be carefully adjusted to get the results shown here.}%
\label{fig2}%
\end{figure}

Fig \ref{fig2} gives a comparison between the true step-wise spectrum and the
Pade analytic continuation results, and the LSF results. The Pade scheme is
not very robust in evaluating the spectral function. One needs to carefully
choose number of Matsubara frequencies to do the analytic continuation before
a reasonable spectrum can be obtained for each temperature. The comparison
seems to show that there is a strong temperature dependence on the calculated
spectrum with the Pade scheme while the dependence is much weaker with the LSF
scheme. It also suggests that both schemes perform better at lower
temperature. This is not surprising as reduced $T$ implies denser Matsubara
frequency points for an analytic continuation to carry out, thus enabling more
information to be used and more features to be extracted out of the original
data. The figure does suggest that at a low temperature the LSF scheme is
out-performed by the Pade scheme. But one might be cautious in coming up with
this conclusion, as quick oscillations could be argued to be an expected
feature of a reasonable analytic continuation
here\cite{comment4_HighOscillation}.%

\begin{figure}[ptb]%
\centering
\includegraphics[
height=4.5446in,
width=6.487in
]%
{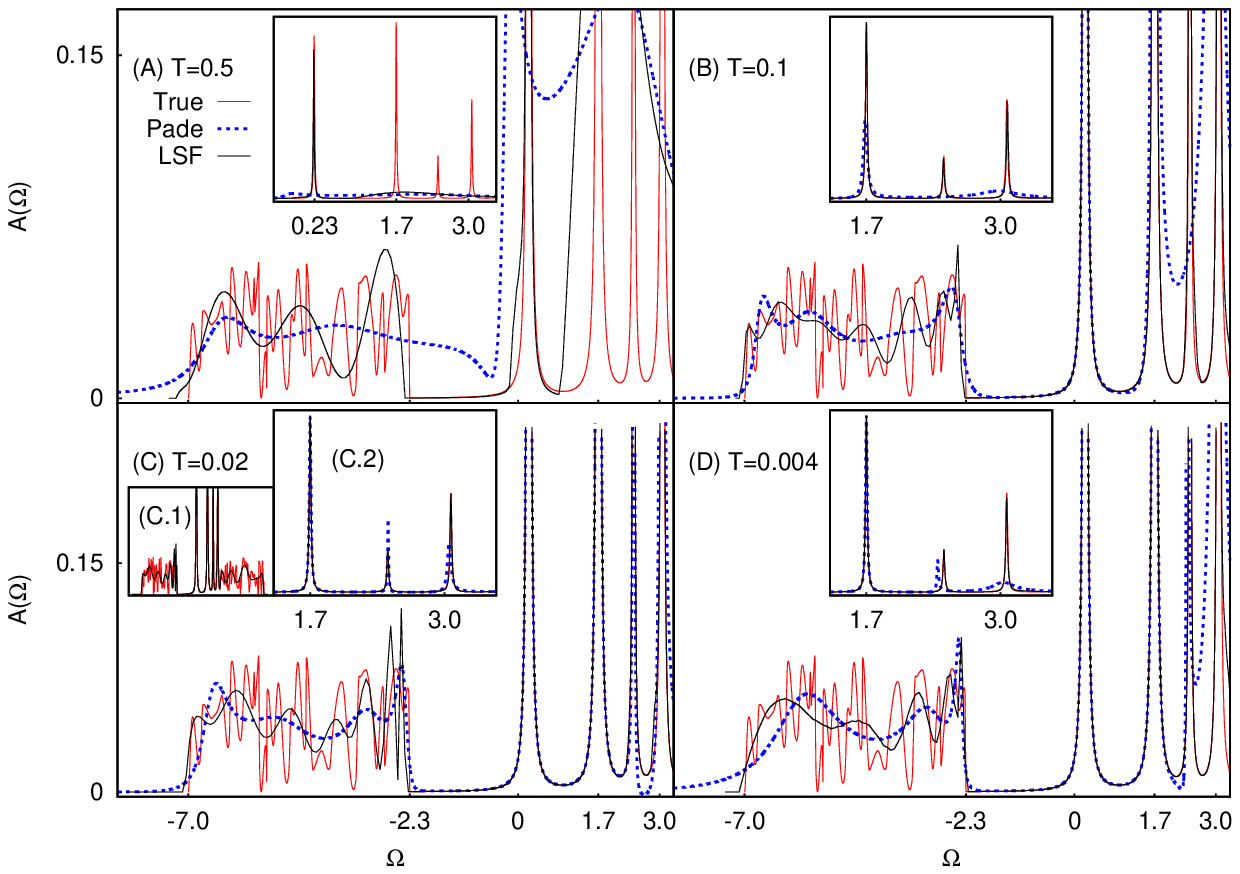}%
\caption{Comparison of the true spectrum composed of two split conduction
bands and four delta peaks with results returned from the Pade and the LSF
schemes under different temperatures. The overall true spectrum, same for all
the four temperature cases, is presented in the left inset of plot (c),
together with the corresponding LSF spectrum as a comparison. In each panel,
only the lower band and the delta peaks are shown for better data
visualization, as the schemes perform alike on both lower and upper bands.
With $\eta$ fixed at $0.01$ for consistent comparison, the number of frequency
points used in the Pade scheme needs to be chosen carefully for $T=0.1$ and
$0.004$ to ensure a valid spectrum function. The inset in each panel presents
an overall view of the delta peaks.}%
\label{fig3}%
\end{figure}

Fig \ref{fig3} presents a comparison between the true spectrum, the Pade
analytic continuation results, as well as the LSF results. The true spectrum
is made up of two conduction bands and four delta peaks. As can be seen in
this specific example, the LSF scheme performs in general better than the Pade
scheme for all the temperatures. At $T=0.5,$ the LSF scheme can nearly
reproduce the delta peak closest to the zero frequency, but misses the other
three peaks which are far away from zero. This is actually a feature of the
LSF scheme. It reveals information best around the zero frequency, thus helps
to identify the spectrum for low energy excitations around the Fermi level of
a physical system. We also see from the plot that the LSF scheme tracks
reasonably well the trend of the varying conduction band, which is also an
advantage of least square based methods due to their smoothing nature. As the
temperature increases, both the Pade and the LSF schemes improve their
performance on reproducing the exact spectral function. The performance is
better at $T=0.02$ where both schemes are able to reproduce the four delta
peaks and track very well the overall trend of the conduction band. But both
schemes do less well at $T=0.004,$ and the Pade scheme suffers more. This can
be ascribed to the limited number of Matsubara frequencies used for the
analytic continuation.

Now let's look at a more practical problem and see how the current scheme
performs on it. To use a relevant model Hamiltonian to describe the physics of
the copper-oxide layer of high temperature
superconductors\cite{Zhang_rice_PhysRevB.37.3759}, we consider the single band
repulsive Hubbard model on a square lattice at half filling,%
\begin{equation}
H=-t\sum_{\left\langle i,j\right\rangle ,\sigma}c_{i,\sigma}^{\dagger
}c_{j,\sigma}+U\sum_{i}n_{i,\uparrow}n_{i,\downarrow} \label{1-4b}%
\end{equation}
with $\left\langle i,j\right\rangle $ denoting the nearest neighbor hopping.
The parameters are chosen at $U=4t$. The bare temperature Green function is
defined as
\begin{equation}
\mathcal{G}_{0}\left(  i\omega_{n},\mathbf{k}\right)  =\frac{1}{i\omega
_{n}-\epsilon_{\mathbf{k}}}%
\end{equation}
where $\epsilon_{\mathbf{k}}=E_{\mathbf{k}}-\mu$ with $E_{\mathbf{k}%
}=-2t\left(  \cos k_{x}+\cos k_{y}\right)  $ and $\mu=U/2$ fixed by the
particle-hole symmetry to ensure half filling. The second order corrected
temperature Green function is calculated as\cite{Beach_PhysRevB.61.5147}
\begin{align}
\mathcal{G}\left(  i\omega_{n},\mathbf{k}\right)   &  =\frac{1}{\mathcal{G}%
_{0}^{-1}\left(  i\omega_{n},\mathbf{k}\right)  -\Sigma\left(  i\omega
_{n},\mathbf{k}\right)  }\\
&  =\frac{1}{i\omega_{n}-E_{\mathbf{k}}-\Sigma^{\left(  2\right)  }\left(
i\omega_{n},\mathbf{k}\right)  } \label{1-4a}%
\end{align}
with
\begin{equation}
\Sigma^{\left(  2\right)  }\left(  i\omega_{n},\mathbf{k}\right)
=-\frac{U^{2}}{N^{4}}\sum_{\mathbf{q},\mathbf{k}^{\prime}}\frac{\left[
f\left(  \epsilon_{\mathbf{k}^{\prime}}\right)  +f\left(  \epsilon
_{\mathbf{q}-\mathbf{k}^{\prime}}\right)  -1\right]  f\left(  \epsilon
_{\mathbf{q}-\mathbf{k}}\right)  -f\left(  \epsilon_{\mathbf{k}^{\prime}%
}\right)  f\left(  \epsilon_{\mathbf{q}-\mathbf{k}^{\prime}}\right)  }%
{i\omega_{n}+\epsilon_{\mathbf{q}-\mathbf{k}}-\epsilon_{\mathbf{k}^{\prime}%
}-\epsilon_{\mathbf{q}-\mathbf{k}^{\prime}}}%
\end{equation}
Here $f\left(  \epsilon_{\mathbf{k}}\right)  $ denotes the usual Fermi-Dirac
distribution function. For the current study, we are concerned about the
spectral function at $\mathbf{k=}\left(  0,0\right)  .$ The true spectrum is
obtained by using Eq. \ref{0-2}. The calculation of $\mathcal{G}\left(
i\omega_{n},\mathbf{k}\right)  $ is performed on a $8\times8,$ as well as a
$100\times100$ lattice, which would result in different spectrum structures
and unknown data errors for analytic continuation schemes to work on.%

\begin{figure}[ptb]%
\centering
\includegraphics[
height=4.0387in,
width=5.7666in
]%
{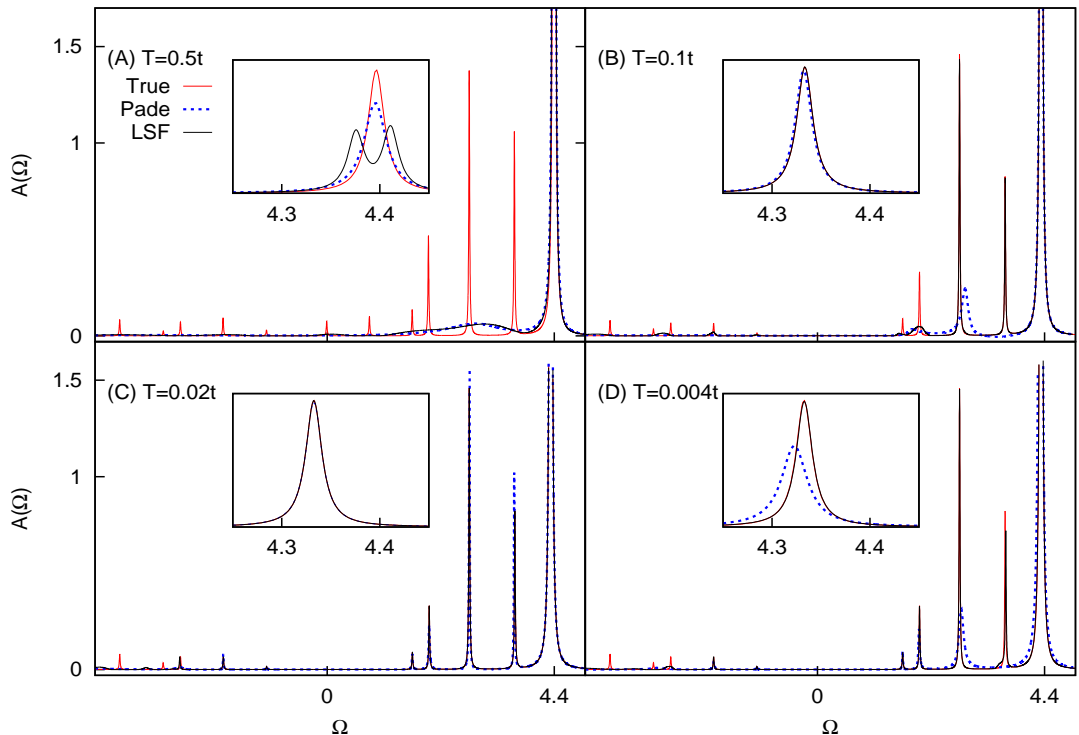}%
\caption{Comparison of the true and the calculated spectral functions using
the Pade and the LSF schemes under different temperatures. Pertinent to the
half-filled Hubbard model in Eq. \ref{1-4b}, the true spectrum is made up of
isolated delta peaks, a natural outcome of calculation done on a small (8 by
8) system. In the insets shows the major delta peak around $\omega\approx4.4$
from different schemes. At $T=0.5t,$ the LSF scheme gives out two delta peaks
differring in their locations by about $0.03$, less than the threshold
difference of $0.09.$ At $T=0.1t,$ the Pade scheme needs to have the number of
Matsubara frequencies adjusted to produce a valid spectral function.}%
\label{fig4}%
\end{figure}

Fig \ref{fig4} shows a comparison of the true spectrum calculated on a 8 by 8
lattice against the results returned by the two analytic continuation schemes
under different temperatures. At $T=0.5t$ in panel A, we see that both schemes
perform similarly: only one peak is reproduced well, all the other peaks are
generated as a continuum. As shown in the inset, the LSF scheme brings up two
delta peaks around $\omega\approx4.4t,$ which, however, could be an artifact
as the peak separation is about $0.03,$ much smaller than the reference width
of $0.09$ argued through Fig \ref{fig1}. For the other temperatures, the LSF
scheme outperforms the Pade scheme in that it gives very accurate peak
locations as well as their intensities. As the temperature is reduced, very
much improved fit is obtained, which illustrates once again the importance of
denser Matsubara frequencies in producing better results. For the Pade scheme,
it performs best at $T=0.02t,$ but still gives wrong intensity of the two
delta peaks located around $\omega\approx3.0t.$ For $T=0.1t,$ the Pade scheme
seems to have serious trouble in producing a valid spectral function with
$\eta$ fixed at $0.01.$ For $T=0.004t,$ the number of Matsubara frequency
points might be too few for a reliable Pade calculation to be carried out, but
this does not seem to affect too much the performance of the LSF scheme.%

\begin{figure}[ptb]%
\centering
\includegraphics[
height=4.0439in,
width=5.7752in
]%
{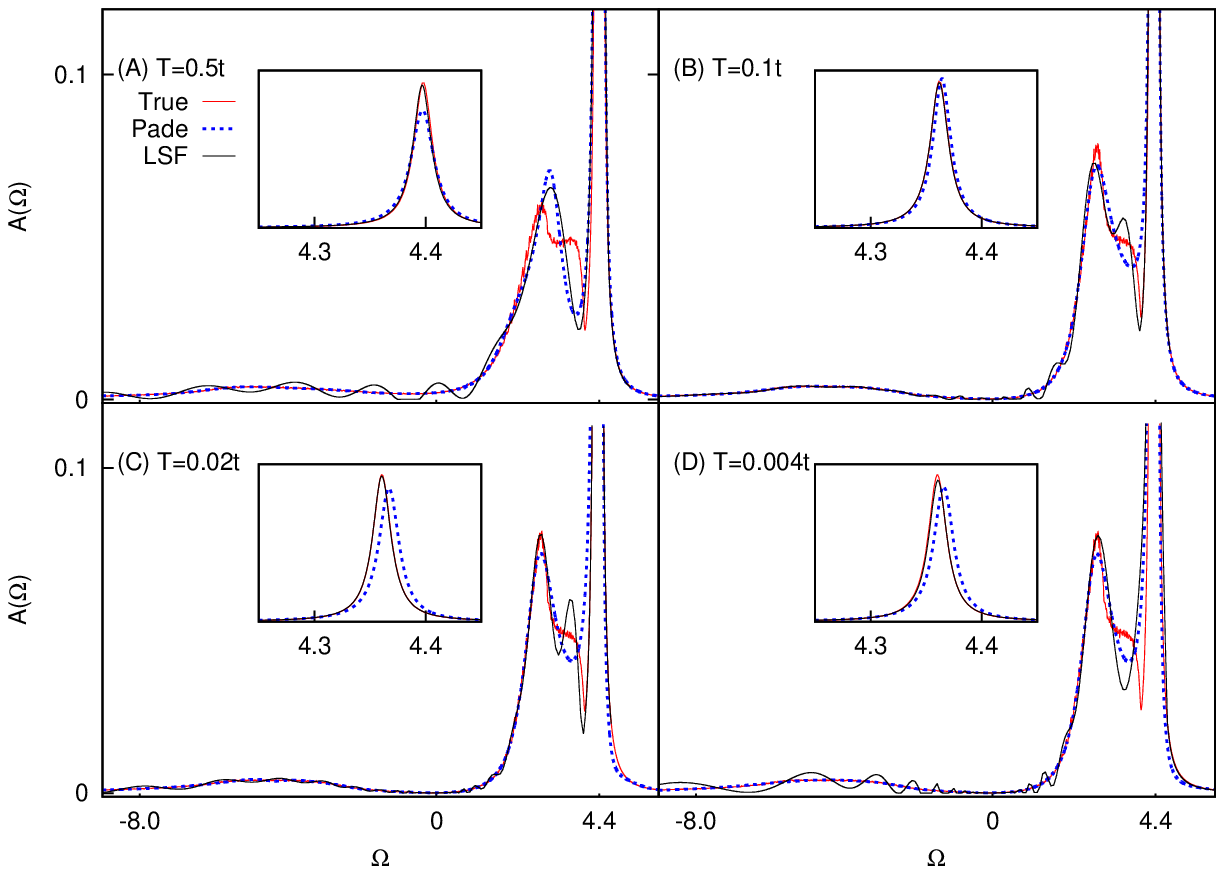}%
\caption{Comparison of the true spectrum of the half-filled Hubbard model on a
100 by 100 lattice against results returned from the Pade and the LSF schemes
under different temperatures. As in Fig \ref{fig4}, subtle details of the
spectral functions are shown in each plot while presented in each inset is the
comparison of the overall behavior of the peak at $\omega\simeq4.4t$ from each
scheme. Due to error cumulation in calculating the input Green function, the
best fit achieveable for $T=0.02t$ and $T=0.004t$ can only reach a residue of
$10^{-14},$ thus the cutoff parameter $\lambda$ is set at $10^{-12}$ for the
calculations of these two cases.}%
\label{fig5}%
\end{figure}

Fig. \ref{fig5}, instead, shows a comparison of the true spectrum on a 100 by
100 lattice against the calculated spectra under different temperatures. For
this specific problem, the Pade scheme does a very good job on reproducing the
broad continuum at negative energy. But the peak at $\omega\approx4.4t$ is not
reproduced equally well. This is quite different for the LSF scheme, where the
major features of the two peaks are reproduced very well, but the continuous
spectrum is superimposed with fluctuations. The performance of the LSF scheme
is most seriously affected by the fluctuations at either too high or too low
temperatures. Further studies are needed to understand why this is the case
and to improve the LSF scheme even further. In the current calculation, the
data precision of the input Green function is identified to be of $10^{-14}$
for $T=0.02t$ and $T=0.004t.$ Thus the Tikhonov cutoff parameter is set at
$\lambda=10^{-12}$ for these two temperatures.

The four examples illustrated above are typical in that they cover cases from
having discontinuity, very rapid change, isolated delta peaks and finite data
precisions in their spectra. From all these cases, we can see that the
performance of the LSF scheme is at least comparable to that of the Pade
scheme but with much weaker temperature dependence. It can be well understood
that both schemes share comparable performance as they both aim at getting
close to $\mathcal{G}\left(  i\omega_{n}\right)  $ in a best possible way. The
only difference is that the LSF scheme preserves the correct $i\omega_{n}$
dependence of $\mathcal{G}\left(  i\omega_{n}\right)  $ while the Pade scheme
approximates $\mathcal{G}\left(  i\omega_{n}\right)  $ with a continued
fraction form. Thus, usually, the LSF scheme gives more accurate extrapolation
for $\mathcal{G}\left(  i\omega_{n}\right)  $ than the Pade scheme if the
fitted data are from small Mastubara frequency points. Due to its smoothing
nature and limited working precision, the LSF scheme cannot give subtle
spectral details but return major features as well as correct trends of hidden
data in a systematic and efficient way. Practically, the LSF scheme only uses
a very limited number of free parameters, such as, the cutoff parameter
pertinant to the Tikhonov regularization, the noise level threshold and the
frequency interval to set up the one dimensional frequency grid. The resulting
spectrum does not critically depend on these adjustable parameters as long as
their values can be reasonably justified.

The current LSF scheme does need further development and testing in several
aspects. The most relevant one is what if delta peaks have a finite broading.
It is still unknown how efficiently the continuous background spectrum
$\tilde{A}\left(  \Omega\right)  $ is able to pick up this feature. A second
concern is in case a delta peak exists within the support of a continuous
spectrum. The fitted delta peak will take up some nearby spectral weight which
necessarily results in a deformation of the continuous background. One more
concern of the current scheme is about the linear approximation of the
continuous spectrum. It would be interesting to see how much one can improve
if the continuous background spectrum is approximated with a cubic
interpolation. All these concerns are either numerically too involved or
impossible to address within the current scheme when the frequency grid is set
up with evenly divided frequency intervals. We reserved them for future work.

In summary, we have shown that the LSF scheme introduced here has comparable
performance aganist the Pade scheme. It is able to systematically uncover
major features of the unknown spectral function hidden in $\mathcal{G}\left(
i\omega_{n}\right)  $. This is made possible by working with a spectral
function and a temperature Green function in the frequency space expressed in
Eq. \ref{0-1} where the high frequency information is suppressed only mildly.
As a comparison, one might be faced with the difficulty caused by the
exponential suppression of a spectral function when the analytic continuation
problem is formulated in the imaginary time
space\cite{early_QMC_chi2_fit_PhysRevB.34.4744}. The analytic continuous
problem is a notoriously difficult one to solve. As is well-known, the
resulting coefficient matrix for the reformulated linear inverse problem is
necessarily ill-conditioned. But, it seems in the current study the Tikhonov
regularization fixes this issue pretty well. Meanwhile, dynamically adding
delta peaks into the linear inverse problem helps to reach the global minimum
of the fitted residue, while using a continuous spectrum background in the fit
helps to improve the efficiency in minimization. As the continuous spectrum is
able to adapt itself to any broad variation, failure to reach the known global
minimum, viz vanishing residue, must be due to sharp features which cannot be
captured by the continuous background. This observation underlies the idea how
to add tentative delta peaks into the fit. As the whole fitting process is
least square based, this necessarily gives out a more robust way to do the
analytic continuation than the Pade scheme. Actually we have tried to argue at
the very beginning of this paper that the Pade scheme is inherently vulnerable
to data errors. So far, the current LSF scheme can only be applied to high
precision data. Given further improvement on its performance and robustness,
this scheme might be used to treat data from a Quantum Monte Carlo simulation,
where the covariance matrix structure is known for the input temperature Green
function. Meanwhile, the current scheme might be easily adapted to other
inverse problems involving integral equations as well. In addition, it might
find much wider applications in both experimental data analysis, say, in
neutron scattering and optical measurements, and many engineering fields if
its performance can be further enhanced. Future efforts on the inverse problem
with the least square fitting scheme will be directed along these dirctions.%

The author would like to acknowledge Junyu Guo, Yi Xue and Shina Tan for many useful discussions.
He would also like to thank Prof. Kai-Ming Ho and Dr. Cai-Zhuang Wang for insightful discussions and kind help.
Research supported by the U.S. Department of Energy, Office of Basic Energy Sciences, Division of Materials Sciences and Engineering.  Ames
Laboratory is operated for the U.S. Department of Energy by Iowa State University under Contract No.  DE-AC02-
07CH11358.%

\bigskip

\bibliographystyle{apsrev4-1}
\bibliography{linear_fit_citation_v2}

\begin{thebibliography}{10}
\bibitem{comment3_SameSpectr4Gt&G}%
  \BibitemOpen
  \bibinfo {note} {Baym and Mermin's theorem implies that a unique spectral
  function exists for a given temperature Green function. Obviously the
  spectral function giving $G$ gives $\tilde{G}$ as well. Thus, there exists no
  other spectral function which gives $\tilde{G}$ except for the one derived
  from $G$.}%
  \bibAnnoteFile{Stop}{comment3_SameSpectr4Gt&G}%
\bibitem{comment1_linear_appr}%
  \BibitemOpen
  \bibinfo {note} {This is actually not a terrible approximation if the
  frequency range of interest is divided into tiny pieces and the continuous
  part of the spectrum is smooth enough.}%
  \bibAnnoteFile{Stop}{comment1_linear_appr}%
\bibitem{comment5_PrevLSF}%
  \BibitemOpen
  \bibinfo {note} {This is a conventional way to approximate an integration.
  But for an inverse problem, this might introduce excessive unvalidated errors
  into the problem as the quality of a solution can be very sensitive to
  errors.}%
  \bibAnnoteFile{Stop}{comment5_PrevLSF}%
\bibitem{comment2_normal}%
  \BibitemOpen
  \bibinfo {note} {We didn't apply the normalization constraint to the current
  calculation. Actually, this constraint was automatically fulfilled to an
  extremely high precision by various spectra of different systems coming out
  of this scheme. This observation gave us additional confidence in this
  method. Specifically, for relatively high temperatures, the normalization
  constraint can be fulfilled to as high as the machine accuracy, while for low
  temperature, the deviation can also reach $10^{-6}$.}%
  \bibAnnoteFile{Stop}{comment2_normal}%
\bibitem{comment4_HighOscillation}%
  \BibitemOpen
  \bibinfo {note} {We can observe this oscillation behavior on the step-wise
  spectral function example treated by the Pade scheme by gradually increasing
  data precision. Formally, we can argue it the following way. We apply an
  inverse Fourier transformation to bring the retarded Green function,
  $G(\omega+i\eta^{+})$, back to the time regime, and we have $\tilde{G}\left(
  t\right) =\tilde{K}\left( t\right) \tilde{A}\left( t\right)$, with
  $\tilde{K}\left( t\right) =\int_{-\infty}^{\infty}\frac{e^{-i\omega t}
  }{\omega+i\eta}d\omega=e^{-\eta t}$ for $t > 0$ and vanishing otherwise. One
  can then solve for $\tilde{A}\left(t\right)$ and transform back to recover
  $A\left(\Omega\right)$. Numerically however, big error will be introduced for
  $\tilde{A}\left(t\right)$ at large $t$ as the kernel $\tilde{K}\left(
  t\right)$ exponentially decays to zero. This leaves unrealistic high
  frequency oscillation behaviors in the final expression of
  $A\left(\Omega\right)$. In making the above arguement, please note
  $\tilde{A}\left(t\right)$ and $\tilde{A}\left(-t\right)$ are closely
  related.}%
  \bibAnnoteFile{Stop}{comment4_HighOscillation}%
\end{thebibliography}

\end{document}